\documentstyle[twoside,fleqn,espcrc2]{article}

\input epsf.sty
\newdimen\psfigsize
\def\psfigure#1 #2 #3 #4 #5{
    \begin{figure}[tbh]
    \vbox{
    \null\vskip-0.2in\hskip#2\epsfxsize=#1 \epsfbox[0 0 350 350]{#4}
    \vskip -0.73in
    \caption {#5 \label{#3}}
    \vskip -0.3in
    \vskip 0.0truein plus0.2truein}
    \end{figure}
}
\newcommand{\AmS}{{\protect\the\textfont2
A\kern-.1667em\lower.5ex\hbox{M}\kern-.125emS}}

\begin{document}

% declarations for front matter
\title{
\vskip -100pt
 \mbox{} \hfill FSU-SCRI-96C-79\\
 \mbox{} \hfill August 1996\\
\vskip  70pt
Critical Higgs Mass and Temperature Dependence
of Gauge Boson Masses in the SU(2) Gauge-Higgs Model}

\author{
F. Karsch
\address{Fakult\"at f\"ur Physik, Universit\"at Bielefeld,
P.O. Box 100131, D-33501 Bielefeld},
T. Neuhaus
\address{Fachbereich Physik, BUGH Universit\"at
Wuppertal, D-42097 Wuppertal, Germany},
A. Patk\'os
\address{Institute of Physics, E\"otv\"os University, Budapest,
Hungary} and
J. Rank
\thanks{Talk presented at the Lattice 96 conference by J.R.}
$^{\mbox{\scriptsize a,\hspace*{-0.5em}}}$
\address{SCRI, Florida State
University, Tallahassee, FL 32306-4052, USA}
}

\begin{abstract}
We study the effective 3-D SU(2) Gauge-Higgs model at finite temperature
for Higgs-masses in the range from $60$ GeV up to $100$ GeV.
The first order
electroweak phase transition weakens with increasing Higgs-mass and
terminates at a critical end-point.
For Higgs-mass values larger than about
$m_{H,c}=75.4(6)$ GeV the thermodynamic signature of the transition
is described by a crossover.
Close to this Higgs-mass value we investigate the vector boson propagator
in Landau gauge. The calculated W-boson screening masses
are compared with predictions based on gap equations.
\end{abstract}

\maketitle

\section{Introduction}

The standard model of electroweak interactions predicts the existence of a
phase transition inbetween a low temperature symmetry broken and a high
temperature symmetric phase \cite{Lin72}.
Its thermodynamic properties lead to
cosmological consequences. One might hope that the
baryon asymmetry can be generated at the electroweak phase transition, if the
transition is of strong first order.
For values
of the zero temperature
Higgs-mass
$m_H \le 70$ GeV the phase transition is of first
order \cite{Fodor,Guertler1}.
As the Higgs-mass is increased further, the thermodynamic singularity
at the electroweak phase transition weakens. It is
even possible, that for large values of the Higgs-mass the phase
transition loses its nonanalytical structure and is described
by a crossover \cite{Kaj}. In the strict sense the electroweak phase
transition would then cease to exist.
The exact determination of the critical Higgs-mass value
$m_{H,c}$, at which the first order electroweak phase transition
changes from first order to a crossover is important in view of its
implications for the standard model.

Another aspect of the electroweak theory is the non-vanishing
magnetic screening mass. It controls the infrared behavior of
the theory at high temperatures and influences the nature of the
phase transition itself.

\section{The 3-D SU(2) Gauge-Higgs model on the lattice}

This work is a continuation of earlier studies and for details
on the formulation of the theory we refer to \cite{Karsch96}.
Here we recall the simulated $3-D$ action functional:
\begin{eqnarray}
&
S^{3D}_{lat}={\beta \over 2}\sum_P{\rm Tr}U_P(x)+{1\over 2} \sum_{x,i}
{\rm Tr}\Phi^\dagger_xU_{x,i}\Phi_{x+\hat{i}} \nonumber \\
&
-{1\over 2\kappa}\sum_x{1\over 2}
{\rm Tr} \Phi^\dagger_x\Phi_x
-{\lambda_3\over 24}\sum_x({1\over 2}{\rm Tr}
\Phi^\dagger_x\Phi_x)^2 .
\end{eqnarray}

The relationship of the dimensionless lattice couplings $\beta,~\lambda_3,~
\kappa$ to the couplings of the $T=0$ $SU(2)$ Gauge-Higgs system
can be found in \cite{Karsch96}. Our simulation is performed at
$\beta=9.0$ and at $5$ values $\lambda_3=0.170190, 0.291275, 0.313860,
0.401087$ and $0.498579$. The hopping parameter is varied across the
phase transition. These $\lambda_3$-values correspond to
Higgs-mass values of $m_H=59.2, 76.1, 78.9, 88.7$
and $98.5$ GeV, if the 1-loop parameter mapping of \cite{Karsch96}
is used. For possible correction terms inducing small corrections
to the here cited Higgs-mass values
we refer to a forthcoming publication.

\section{Determination of the critical Higgs-mass}

To illustrate the critical behavior as a function of $\lambda_3$, or the
Higgs-mass, we display
in figure \ref{cv_max} the maxima of the
$\Phi^2$-susceptibility, which
is given by
\begin{equation}
\chi_{\Phi^2} = V  \left\langle \left( \Phi^{\dagger}\Phi -
\langle \Phi^{\dagger}\Phi \rangle \right)^2 \right\rangle .
\label{chi}
\end{equation}

\psfigure 3.0in -0.5in {cv_max} {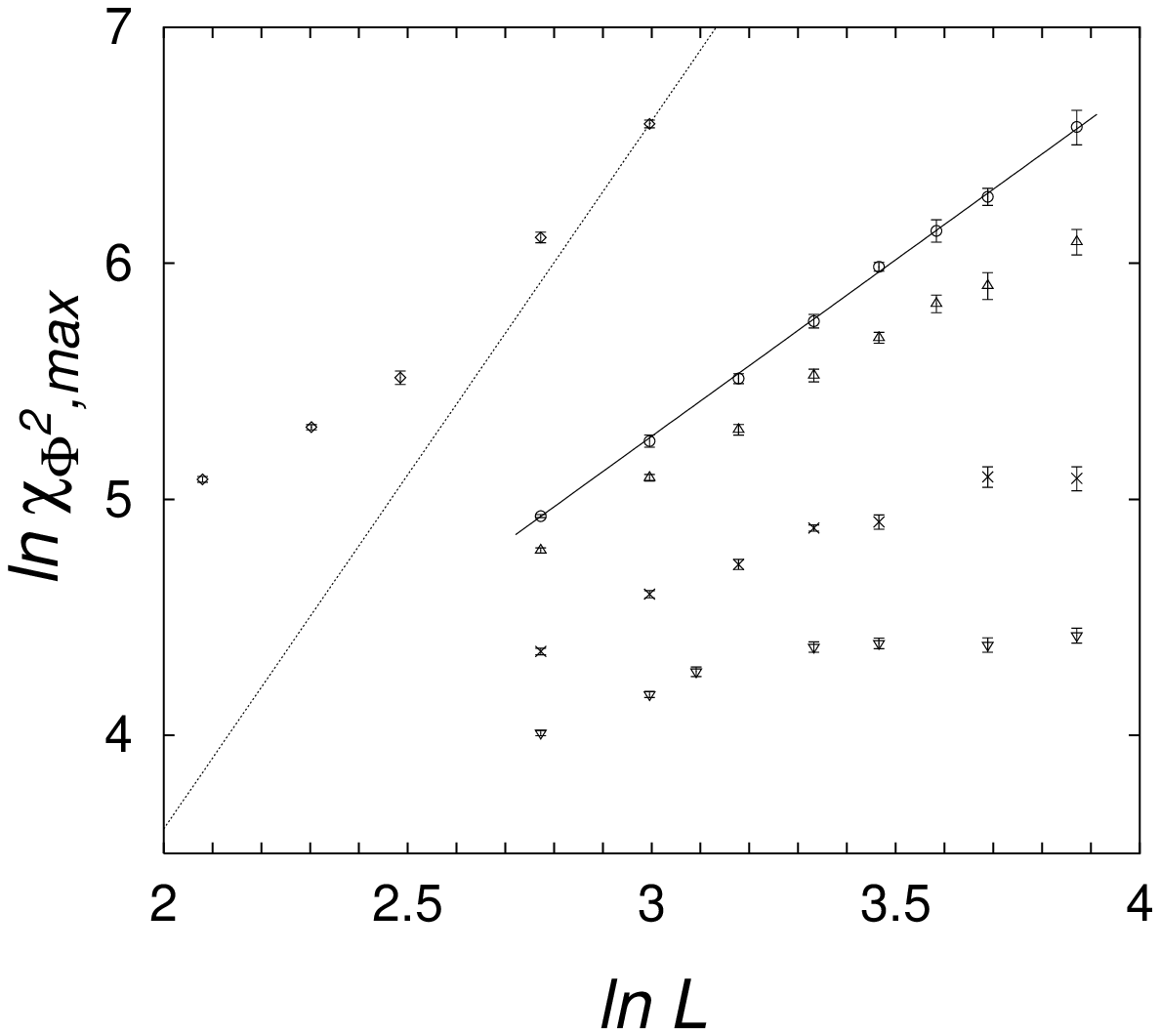} {Maxima of the
$\Phi^2$-susceptibility.}

We investigate lattice sizes ranging from $L=8$ up to
$L=48$. The labeling of the data in figure \ref{cv_max} corresponds to the one
used in figure \ref{l_im}.
The data at
$\lambda_3=0.170190$ are consistent with a first order transition, the
dotted line with slope $3$ in the figure. At
$\lambda_3=0.313860,
0.401087$ and $0.498579$ a crossover is observed, while
at $\lambda_3=0.291275$ the behavior is almost critical,
the solid scaling curve in figure \ref{chi}.

The determination of the critical Higgs-mass value relies
on the analysis of Fisher or Lee-Yang zeroes
\cite{Lee-Yang} in the crossover-region of the theory.
The partition function $\cal{Z}$ is analytically continued
into the complex plane as a function of the complex hopping parameter
$\kappa$. Denoting with $z_0$ the lowest zero of $\cal{Z}$, i.e.
the zero in $\kappa$ with smallest length,
we expect in the vicinity of the critical end-point
the scaling law
\begin{equation}
\mbox{Im}(z_0) = C  L^{-1/{\tilde{\nu}}}+R(\lambda_3) .
\label{im_kappa}
\end{equation}
Such a scaling behavior can also be observed for  Lee-Yang
zeroes in the high temperature phase of the Ising model.
Our strategy to localize the end-point then is to
determine the value of $\lambda_3$, at which the regular
contribution $R$ to the scaling law vanishes: $R(\lambda_{3,c})=0$.
In figure \ref{l_im} we display $\mbox{lnIm}(z_0)$ vs. $\mbox{lnL}$. The solid
curves in figure \ref{l_im} correspond to fits with the scaling law.

\psfigure 3.0in -0.5in {l_im} {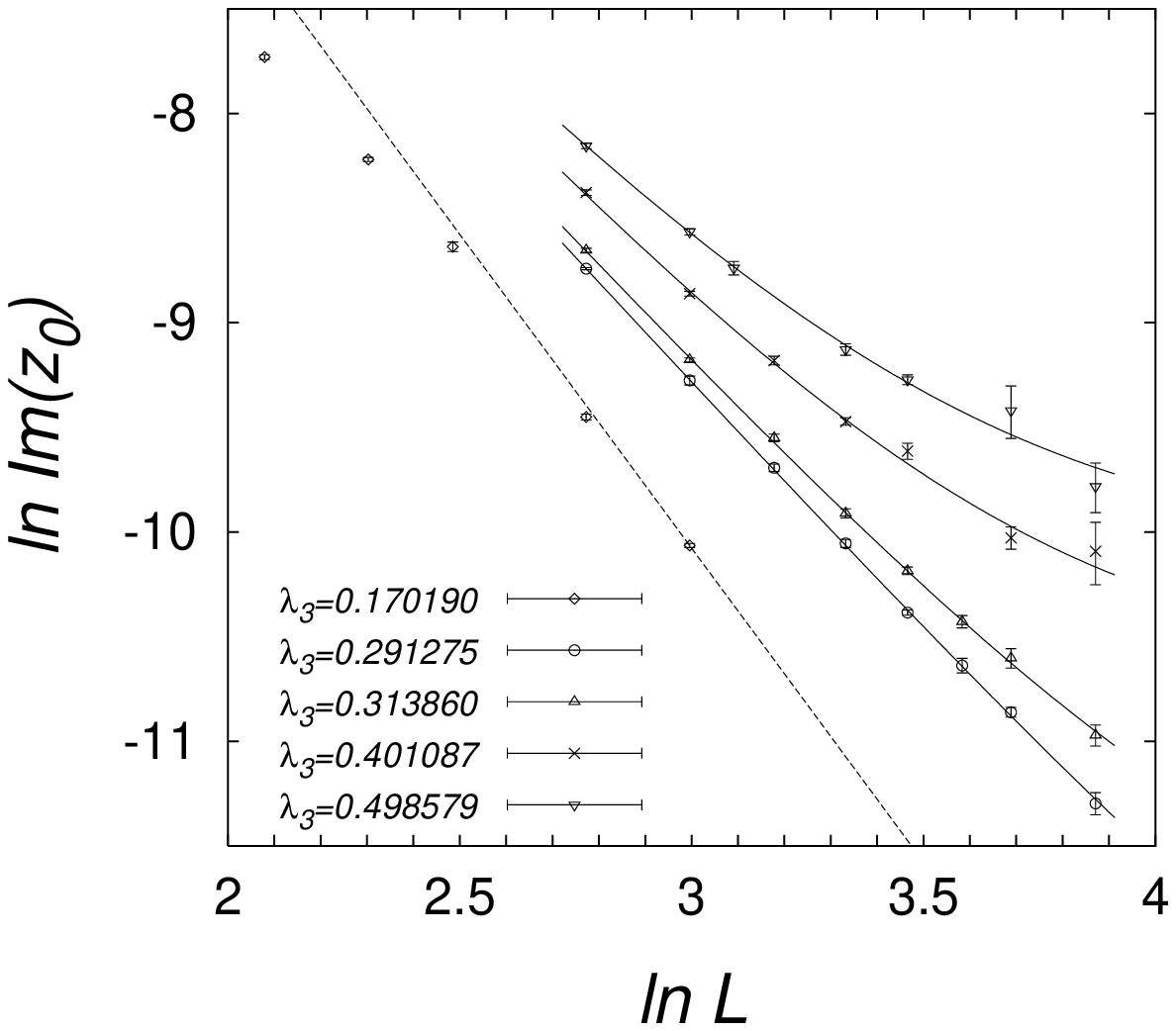} {Imaginary parts of the lowest
zeroes of the partition function.}

In figure \ref{fitconst} the constant $R(\lambda_3)$ is displayed
as function of $\lambda_3$. The data
are consistent with a linear dependence on $\lambda_3$ and the
fit results into $\lambda_{3,c}=0.2853(48)$, corresponding to
a critical Higgs-mass value of approximately $m_{H,c} = 75.4(6)$ GeV.

\psfigure 3.0in -0.5in {fitconst} {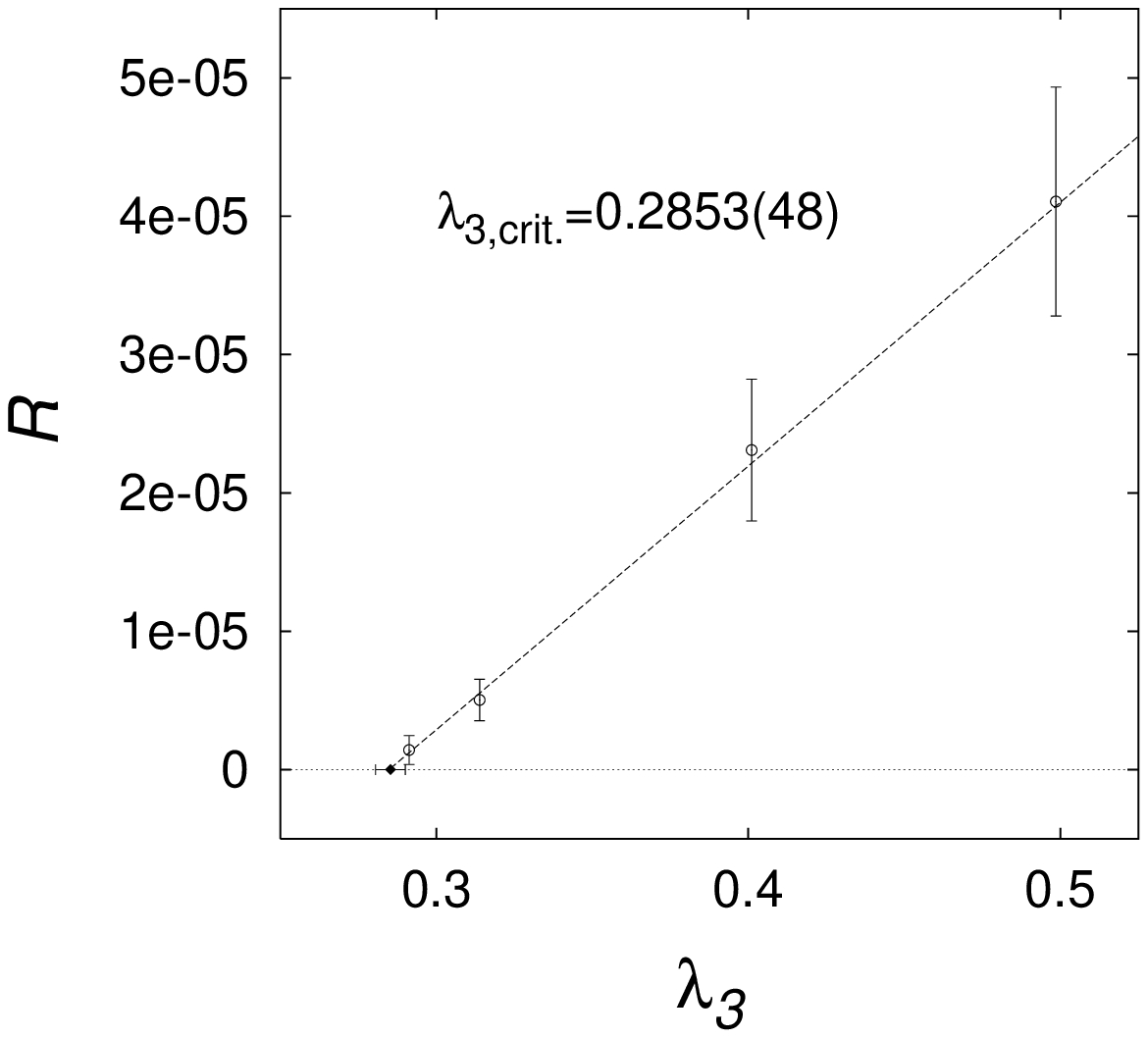} {The determination of
$\lambda_{3,c}$.}

\section{Gauge Boson masses in Landau gauge}

The W-boson mass is determined
from the W-boson propagator
in Landau-gauge, $|\partial^{\mu} A_{\mu}(x)|^2=0$. For a detailed discussion
of the implementation of the Landau-gauge on a lattice see
\cite{Karsch96} and references therein.
We investigate the W-boson propagator at
$\lambda_3=0.291275$, which is
very close to the critical Higgs-mass.
Simulations have been performed on a $16^2 \times 32$ lattice. The
propagators
have been analyzed in the same way as discussed in \cite{Karsch96}. Our
results are shown in figure \ref{mass_w}.
\psfigure 3.0in -0.6in {mass_w} {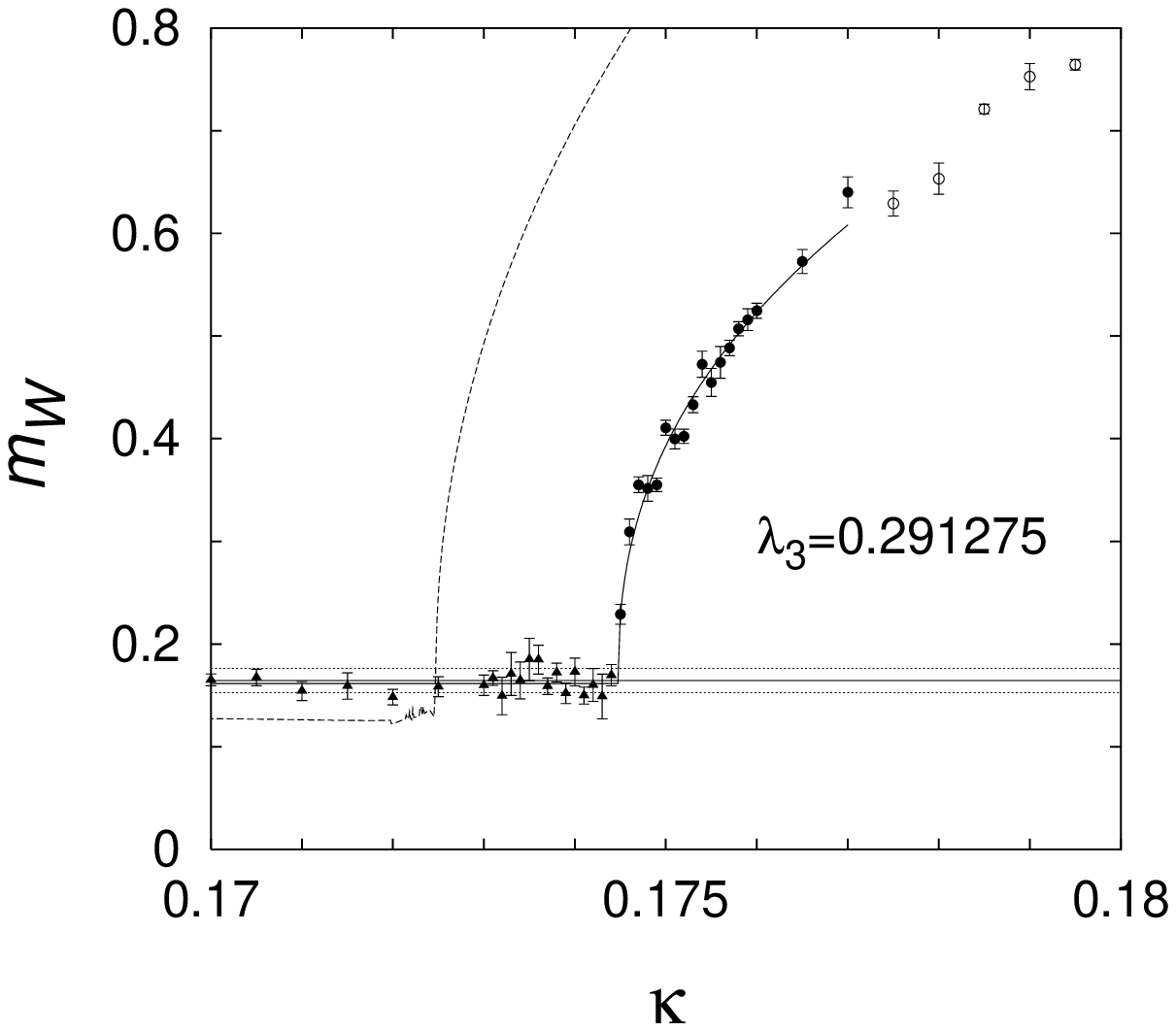} {W-boson screening masses,
calculated on a $16^2 \times 32$ lattice at
$\lambda_3=0.291275$.
The full curve describes the
fit to the data. The dashed curve represents the result obtained from
gap equations.}

In the high temperature phase the W-boson propagator stays
constant. A fit to the data for $\kappa \le \kappa_c$, the full
triangles in the figure, yields
$m_W(\kappa \le \kappa_c) = 0.161(3)$.
It is worthwhile to note, that in
pure SU(2) gauge theory the magnetic screening mass has a value $0.165(12)$,
the horizontal full and dotted lines in figure \ref{mass_w}.
In the symmetry broken phase the mass increases rapidly. In this
region the data are very well described by the ansatz
\begin{equation}
m_w = 0.161 + a (\kappa - \kappa_c)^\beta \quad \kappa \ge \kappa_c~.
\label{mwfita}
\end{equation}
The fit to the full circles in the figure results into a value
for the exponent $\beta$ of $\beta \approx 0.4$.
We are now able to compare our results with predictions based on gap
equations \cite{Buc95}. Using similar parameters in the gap equations
as in our study we observe a qualitative
agreement, the dashed curve in the figure.

\section{Summary}
We have exploited the scaling behavior of partition function zeroes
in the vicinity of the critical end-point in order to
determine
the
critical Higgs-mass $m_{H,c}$. The electroweak phase transition
looses its first order character at a Higgs-mass
value of about $m_{H,c} = 75.4(6)$ GeV.
Close to the critical Higgs-mass we have measured the W-boson propagator in
Landau gauge. W-boson screening masses remain
constant in the high temperature symmetric phase and increase
in the low temperature Higgs phase.
The agreement of these data with predictions based on gap equations is
of
qualitative nature.

\end{document}